\documentclass[a4paper,aps,prl,twocolumn,longbibliography,english,superscriptaddress]{revtex4-2}

\usepackage{amssymb}
\usepackage{graphicx}
\usepackage{dcolumn}
\usepackage{bm}
\usepackage{amsmath}
\usepackage{subfigure}
\usepackage{float}
\usepackage{color}
\DeclareUnicodeCharacter{202C}{ }
\usepackage[colorlinks=true,	linkcolor=blue,urlcolor=blue,anchorcolor=blue,citecolor=blue,bookmarksnumbered]{hyperref}

\begin{document}


\title{Suppression of motional dephasing using state mapping}

\author{Yuechun Jiao}
\thanks{These authors contributed equally to this work.}
\affiliation{State Key Laboratory of Quantum Optics Technologies and Devices, Institute of Laser Spectroscopy, Shanxi University, Taiyuan 030006, China}
\affiliation{Collaborative Innovation Center of Extreme Optics, Shanxi University, Taiyuan 030006, China}

\author{Changcheng Li}
\thanks{These authors contributed equally to this work.}
\affiliation{State Key Laboratory of Quantum Optics Technologies and Devices, Institute of Laser Spectroscopy, Shanxi University, Taiyuan 030006, China}

\author{Xiao-Feng Shi}
\thanks{These authors contributed equally to this work.}
\email{xshi@hainanu.edu.cn}
\affiliation{Center for Theoretical Physics and School of Physics and Optoelectronic Engineering, Hainan University, Haikou 570228, China}
\affiliation{School of Physics, Xidian University, Xi’an, 710071, China}

\author{Jiabei Fan}
\affiliation{State Key Laboratory of Quantum Optics Technologies and Devices, Institute of Laser Spectroscopy, Shanxi University, Taiyuan 030006, China}

\author{Jingxu Bai}
\affiliation{State Key Laboratory of Quantum Optics Technologies and Devices, Institute of Laser Spectroscopy, Shanxi University, Taiyuan 030006, China}
\affiliation{Collaborative Innovation Center of Extreme Optics, Shanxi University, Taiyuan 030006, China}

\author{Suotang Jia}%
\affiliation{State Key Laboratory of Quantum Optics Technologies and Devices, Institute of Laser Spectroscopy, Shanxi University, Taiyuan 030006, China}
\affiliation{Collaborative Innovation Center of Extreme Optics, Shanxi University, Taiyuan 030006, China}

\author{Jianming Zhao}%
\email{zhaojm@sxu.edu.cn}
\affiliation{State Key Laboratory of Quantum Optics Technologies and Devices, Institute of Laser Spectroscopy, Shanxi University, Taiyuan 030006, China}
\affiliation{Collaborative Innovation Center of Extreme Optics, Shanxi University, Taiyuan 030006, China}

\author{C. Stuart Adams}
\email{c.s.adams@durham.ac.uk}
\affiliation{Joint Quantum Centre (JQC) Durham-Newcastle, Department of Physics, Durham University, DH1 3LE, United Kingdom}

\date{\today}
\begin{abstract}
Rydberg-mediated quantum optics is a useful route toward deterministic quantum information processing based on single photons and quantum networks, but is bottlenecked by the fast motional dephasing of Rydberg atoms. Here, we propose and experimentally demonstrate suppressing the motional dephasing by creating an {\it a priori} unknown but correct phase to each Rydberg atom in an atomic ensemble. The phase created is exactly proportional to the unknown velocity of the thermal motion, resulting in a condition as if no thermal motion occurs to the Rydberg atom upon the retrieval of the signal photon. Our experiments, though hampered by the noise of lasers and the environment, demonstrate more than one order of magnitude enhancement of the coherence time. The feasibility of realizing long-lived storage of single photons in strongly interacting Rydberg media sheds new light on Rydberg-mediated quantum nonlinear optics.

\end{abstract}
\maketitle

Maintaining the coherence in quantum systems is interesting in both fundamental physics \cite{Brune1996,Minev2019} and quantum information processing~\cite{ofek016,scholl2023, gupta2024, Bluvstein2024, dasilva2024a, saei2024quantum}. 
In particular, suppressing the dephasing caused by thermal fluctuations in quantum systems can potentially enable functional quantum devices~\cite{Du2009}. Techniques to reduce motional dephasing of quantum superpositions include spin echo \cite{Hahn1950, rui2015a} and bang-bang \cite{Viola1998,Viola1999,Gotz2007,Alonso2016}. A collective quantum superposition state known as a Rydberg polariton~\cite{firstenberg2016a} is potentially important in the context of single photon sources \cite{dudin2012Strongly,peyronel2012Quantumb,Maxwell2013}, optical transistor \cite{Hofferberth2014,tiarks2014SinglePhoton,tiarks2019Photon}, all-optical quantum gates \cite{Ourjoumtsev2022,Durr2022} and fast read-out of quantum information~\cite{spong2021collectively}. However, progress in Rydberg polariton quantum technology has been hindered by fast motional dephasing on which no effective methods exist for undoing it. 

\begin{figure*}
    \centering
    \includegraphics[width=\linewidth]{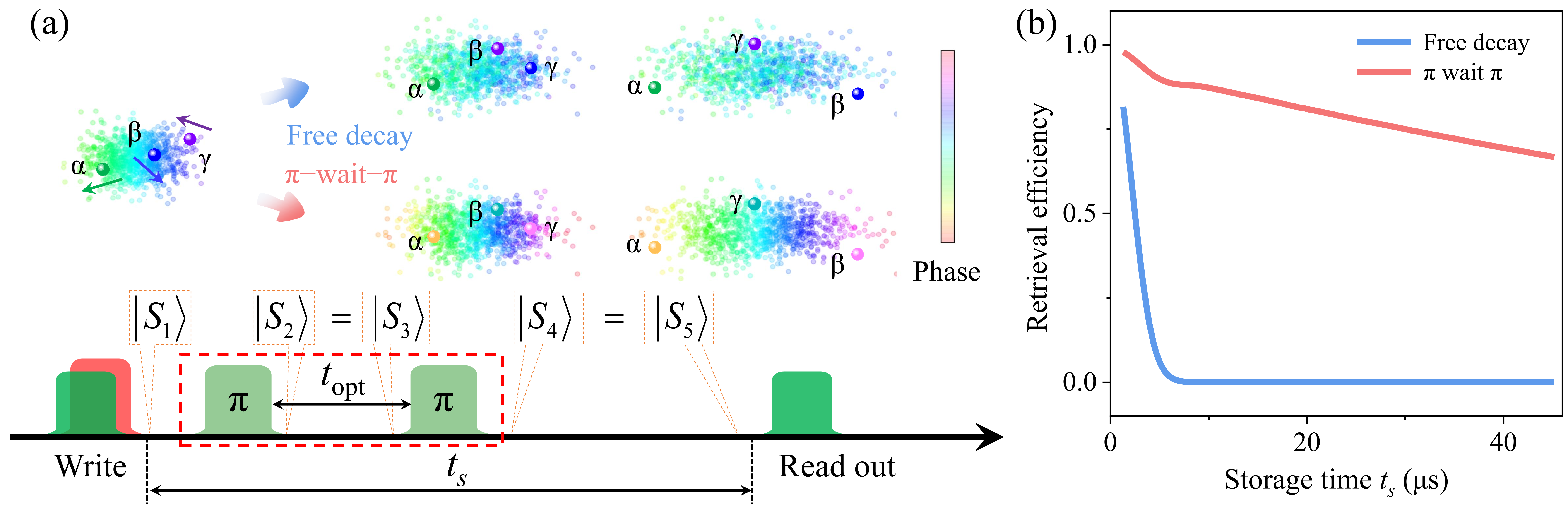}
    \caption{Suppression of the motional dephasing. The upper panel of (a) shows that upon loading the photon, the Rydberg wavefunction in each atom has a certain color representing the $z$-dependent phase corresponding to its initial position $z_j(0)$. No phase change occurs to the atomic state, but the position of the atoms changes due to their thermal motion, as indicated by the arrows. For a free decay, at a later time, the phase of each Rydberg atom does not match the desired phase, resulting in the dephasing of RWS. The associated loss of retrieval efficiency is shown by the blue curve in (b). For the $\pi-$wait$-\pi$ protocol the phase of the Rydberg atom is changed via state mapping. The lower panel in (a) shows the time sequence of the protocol. After a storage time $t_s$ determined by an optimal wait time $t_\emph{w}=t_{\rm{opt}}$, the phase of each atom is exactly equal to the desired value resulting in high efficiency read-out of the photon, shown as the red curve in (b), where the drop of the efficiency is nearly only from the Rydberg-state decay.
    For (b) the parameters used 
in our experiments are shown below Eq.~S16.  }
    \label{Fig.1}
\end{figure*}

For Rydberg polariton, by writing a single photon into an ultracold atomic gas, one obtains the collective state
\begin{eqnarray}
|S_1\rangle = \frac{1}{\sqrt{N}}\sum_{j=1}^{N}e^{ikz_j(0)} |gg\cdots r_1^{(j)}ggg\rangle,\label{Wstate}
\end{eqnarray}
where $k$ is the effective wavevector of the excitation lasers, $z_j(0)$ is the initial position of atom $j$, $\lvert g\rangle$ denotes an atom in the ground state and $\lvert r_1^{(j)}\rangle$ denotes that atom-$j$ is in an excited state $\lvert r_1\rangle$. Owing to their utility in quantum technology~\cite{Adams2020}, we shall assume that $\lvert r_1\rangle$ is a highly-excited Rydberg state, and refers to this collective state as the Rydberg W state~(RWS). If atoms were frozen in space, then at an arbitrary time $t>0$, a coherent retrieval of the signal photon can be realized. Unfortunately, each atom $j$ in the gas undergoes random thermal motion, so that the actual location $z_j(t)$ is no longer equal to $z_j(0)$. Consequently, the phase in the collective state to $kz_j(0)$, does not match the required phase $k(z_j(0)+v_jt)$ needed to read out the photon at a time $t>0$ [see the free-decay illustration of Fig.~\ref{Fig.1}(a)]. This leads to a degradation of the read-out efficiency at time $t$ [see the blue curve of Fig.~\ref{Fig.1}(b) or section II of the Supplemental Material (SM)~\cite{SM} for more details]. If there were an analogous dynamical decoupling protocol as spin echo~\cite{Hahn1950} or quantum bang-bang~\cite{Viola1998,Viola1999,Gotz2007,Alonso2016}, i.e., if one were able to devise a similar control sequence to let each of the atoms reverse its specific velocity, then it would be possible to have the atoms return to their original position at a certain time. In other words, if we were able to have $z_j(t)=z_j(0)$ at a certain `echo' time $t>0$, then we could recover a coherence revival, and thereby quench the motion-induced dephasing. Unfortunately, there is no global control sequence that can reverse the real-space thermal motion of each atom. Consequently, motional dephasing of RWS has proved to be a major stumbling block for the advancement of RWS-based quantum technology, and it looks extremely difficult to remove it~\cite{kurzyna2024}. Protocols‬ have‬ been‬ proposed to‬ reduce‬ motional dephasing‬. For example, in \cite{finkelstein2021}, a scheme that relies on a small value of $k$ was demonstrated. However, this is challenging to implement for RWSs because the wavevector is large. A scheme to remove phases linear in the atomic velocity was proposed in~\cite{shiprap2020}. However, we still need a scheme to create the desired linear phases.
 
Here, we develop and implement a novel protocol, $\pi-$wait$-\pi$, that can overcome these challenges. Instead of returning a dephased quantum superposition back to its original state as in~\cite{Hahn1950,Viola1998,Viola1999,Gotz2007,Alonso2016}, we correct the phase error by creating an {\it a priori} unknown phase that is appropriate at the onset of retrieval. This is distinct from the previous proposal that removes the velocity-dependent phase~\cite{shiprap2020}. 
 The comparison between our approach and previous dynamical decoupling theory is highlighted in section I of the SM. In particular, we devise a scheme to exploit the phase of external control lasers, as displayed in Fig.~1(a).
 The protocol is illustrated as follows: 
 (i) The random atomic motion of  atom $j$ with velocity $v_j$ causes a spurious phase on the Rydberg state component that produces an error at the time of retrieval, $t_s$. Dephasing occurs because of the different $v_j t_s$ for each atom. To remove the spurious phase, we use a state mapping between nearby Rydberg states in the RWS so as to imprint a phase that should exactly cancel the spurious phase. To achieve this for each atom $j$, we apply a $\pi$ pulse to map $\lvert r_1^{(j)}\rangle$ to $\lvert r_2^{(j)}\rangle$, converting the RWS in Eq.~(\ref{Wstate}) to $\lvert S_2\rangle$. This imprints a phase change $\phi_1 = -\pi/2-k_{r}v_j\pi/2\Omega_{r}$ on the Rydberg component in each atom, where $k_r$ is the wavevector of the rephasing laser; for brevity, hereafter $z_j(0)$ in Eq.~(\ref{Wstate}) is absorbed in the definition of the initial state. (ii) We then let the atoms move freely for an `optimal' wait time, $t_{\rm{opt}}$. The atom $j$ moves a distance $v_jt_{\rm{opt}}$, but its internal state remains, so the RWS at the end of the wait, $\lvert S_3\rangle$, shown in Fig.~\ref{Fig.1}(a), is the same as $\lvert S_2\rangle$. (iii) Subsequently, we apply a second $\pi$ pulse to restore the Rydberg state back to $\lvert r_1\rangle$, during which another phase change $\phi_2 = -\pi/2+k_{r}v_j\pi/(2\Omega_{r} )+k_{r}v_j(t_{\text{opt}}+\pi/\Omega_{r})$ appears in the Rydberg component for each atom. 
 Hence the net effect of the two $\pi$ pulses is a phase change $\phi=\phi_1 +\phi_2 =k_{r}v_j(t_{\text{opt}}+\pi/\Omega_{r})-\pi$. With an optimal wait duration $t_{\text{opt}}=k t_{s}/k_{r} - \pi/\Omega_{r}$, we have $\phi=kv_j t_{\text{s}}-\pi$, where the phase $-\pi$ applies to all atoms and therefore is trivial. (iv) Finally, we retrieve the signal at time $t=t_{\text{s}}$ such that the phase imprinted on atom $j$ is the desired one since $v_j t_{\text{s}}$ is exactly the distance atom $j$ travels from $t=0$ to $t=t_{\text{s}}$. In other words, it is as if the RWS just forms upon the retrieval of the signal photon, and the motional dephasing disappears, leaving Rydberg-state decay as the only dephasing channel. The theoretical rephasing effect with parameters used in our experiments is shown by the red curve in Fig.~\ref{Fig.1}(b), where the decrease of the retrieval efficiency is from the radiation decay of Rydberg states. See Section III of the SM for details.
 
 The protocol depends on the condition that the 
light shifts of the rephasing lasers are larger than the Doppler detuning, $\Omega_{r}\gg kv$~\cite{shiprap2020}. For typical experimental parameters pertaining to a laser-cooled atomic ensemble at $40~\mu$K this condition is satisfied even at modest laser powers with $\Omega_{r}\sim2\pi\times1$~MHz. It is also crucial that $v_j$ remains constant for each atom $j$ during the process for otherwise the created phase $\phi$ can't compensate the spurious phase. In practice, $v_j$ can change because of acceleration due to gravity, the trapping potential, and collision between atoms, but these effects are negligible in our case~\cite{schmidt-eberle2020Darktime}.

\begin{figure}
    \centering    
    \includegraphics[width=\linewidth]{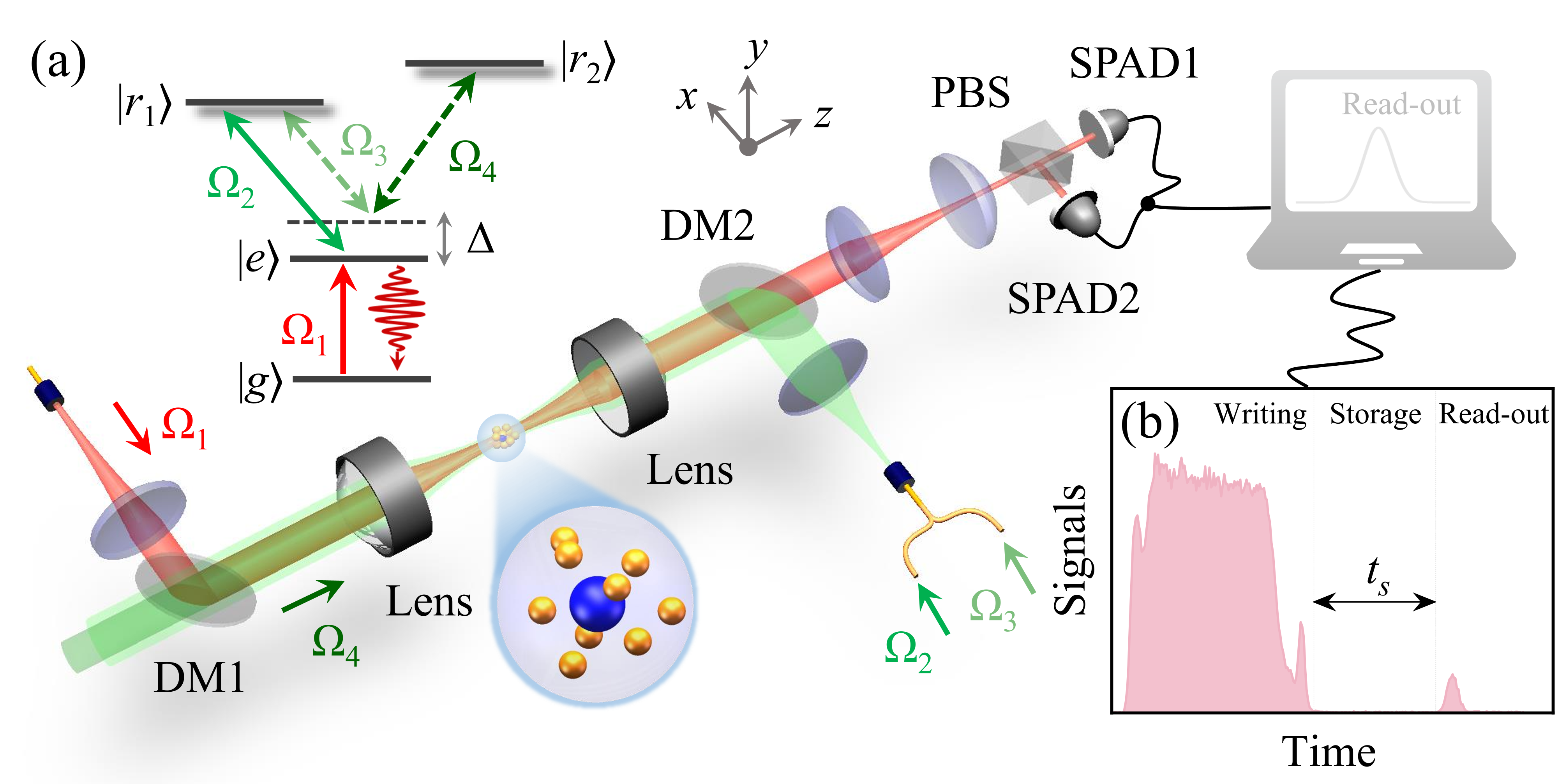}    \caption{Experimental realization. (a) Schematic of the experimental setup and relevant energy levels. The signal photon is stored in a Rydberg state $\vert r_1\rangle$ in the form of RWS in a cold atomic ensemble using Rydberg EIT. The fields denoted by Rabi frequency $\Omega_1$ (852~nm) and $\Omega_2$ (509~nm) resonantly couple the two-photon transition $|g\rangle \rightarrow |r_1\rangle$ via the intermediate state $|e\rangle$. Two rephasing lasers ($\Omega_3$ and $\Omega_4$) drive the transition between $|r_1\rangle$ and $|r_2\rangle$ via $|e\rangle$ with a large detuning $\Delta$. (b) The intensity measured during the complete sequence: Writing, Storage ($t_s$) and Read-out.}
    \label{Fig.2}
\end{figure}

In the case of a RWS, our novel phase correction scheme has two particular strengths: (i) It preserves the Rydberg character of the RWS throughout such that properties such as strong interaction and Rydberg blockade are maintained. This means that there is negligible probability to create more than one Rydberg excitation that could lead to extra many-body dephasing~\cite{dudin2012Strongly}; (ii) It works using an optimal laser configuration, i.e., when the signal and coupling lasers counter-propagate in the ladder-type configuration so that the smallest $\mathbf{k}$ appears in the RWS. In this case, the residual motional dephasing during the loading and retrieval stage, which, unfortunately, can't be removed, is minimal. 

The experimental setup and time sequence are shown in Fig.~\ref{Fig.2}(a) and bottom of Fig.~\ref{Fig.1}(a), respectively. A quasi-one dimensional ensemble of Cs atoms was captured by an optical dipole trap created by a 1064~nm laser with the $1/e^2$  waist $\emph{w}_1 = 4.3~\mu$m. A signal pulse 
($\Omega_1$) of duration 650~ns with a mean photon number of about 1.0 is focused into the ensemble with a $1/e^2$  waist of $\emph{w}_p = 2.5~\mu$m. The signal photon is resonant with the transition $|g\rangle=|6S_{1/2}, F=4, m_{F}=4\rangle\rightarrow|e\rangle=|6P_{3/2}, F'=5, m_{F'}=5\rangle$. A strong coupling laser 
($\Omega_2$) with a beam waist $\emph{w}_c = 8.5~\rm{\mu}$m drives the transition $|e\rangle \rightarrow |r_1\rangle=|65S_{1/2}\rangle$. 
The signal photon is converted into an RWS via the coupling laser. The signal and coupling lasers have opposite circular polarizations and counter-propagate through the atomic ensemble. To store the signal photon, the intensity of the coupling laser is ramped down to zero before the signal laser is switched off. Due to the Rydberg blockade, only one polariton can be stored~\cite{li2016Quantuma,Adams2020}. During the storage time $t_s$, we implement the $\pi-$wait$-\pi$ scheme shown in Fig.~\ref{Fig.1}(a) via a pair of Raman (rephasing) lasers
($\Omega_3$ and $\Omega_4$) to drive the transition $|r_1\rangle \leftrightarrow |r_2\rangle=|70S_{1/2}\rangle$ via $|e\rangle$ with a detuning $\Delta$ = 2$\pi \times$ 335~MHz. Note that the two rephasing lasers counter-propagate and the $\Omega_3$ field co-propagates with the coupling laser [see section III of the SM]. At the end of the storage of duration $t_s$, we read out the RWS and the retrieved photon is detected by a single-photon detector. Figure~\ref{Fig.2}(b) shows the measured transmission of the signal laser during the sequence.

\begin{figure}[htbp]
    \centering
    \includegraphics[width=\linewidth]{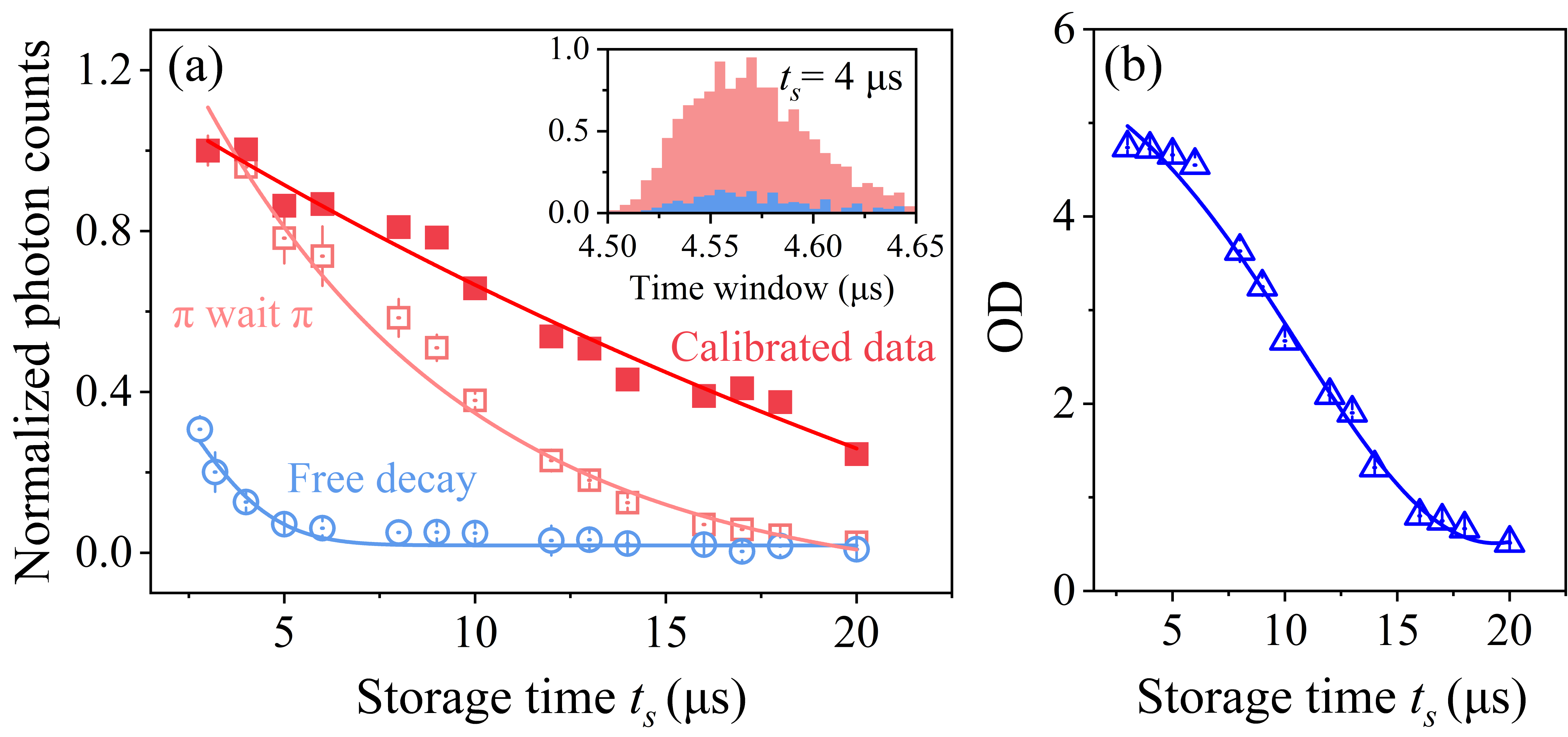}
    \caption{ Long-time storage based on the $\pi-$wait$-\pi$ protocol. (a) Comparison of photon retrieval signals for free decay (circles) and $\pi-$wait$-\pi$ protocol (hollow squares). The solid squares present the corrected retrieval signal by taking the decreasing of OD into account with the fitting line in (b).
    The solid lines are the corresponding fittings. All data at storage time $t_s$ are normalized to the first data point of the $\pi-$wait$-\pi$ protocol at $t_s$ = 2.79~$\mu$s. The error bars show the standard deviation of three independent measurements. Inset: Statistical distribution of retrieval photon counts for free decay (blue area) and the $\pi-$wait$-\pi$ protocol (red area) at storage time of 4~$\mu$s. (b) Measurements of OD for different storage times and the fitting curve with a Gaussian function.}
    \label{Fig.3}
\end{figure}

To experimentally verify the validity of our theory, we measure the number of retrieval photons, integrated over the retrieval time window, as a function of storage time $t_s$. The data are normalized to the first data point of the
$\pi-$wait$-\pi$ protocol, shown in Fig.~\ref{Fig.3}(a). The wait time $t_\emph{w}$ between the two $\pi$ pulses is equal to a $t_s$-dependent 
optimal duration $t_{\rm{opt}}$, and the effective Rabi frequency of rephasing lasers is $\Omega_r = 2\pi \times 1$~MHz with $\Omega_{3(4)} = 2\pi \times$ 21(32)~MHz. The retrieval photon counts are almost absent with a storage time beyond 4~$\mu$s (circles) for the free decay case, 
while shoot up dramatically with the $\pi-$wait$-\pi$ protocol (hollow squares), resulting in a survival time approaching about 20~$\mu$s. The solid lines are fittings with $A_1+B_1\exp{-(t_s/\tau_1)^2}$ and $A_{2}+B_{2}\exp{-(t_s/\tau_{2})}$ for the free decay and the $\pi-$wait$-\pi$ protocol, yielding $\tau_1=3.29~\mu$s and $\tau_2=7.06~\mu$s, respectively. The inset of Fig.~\ref{Fig.3}(a) shows the statistical distribution of individual retrieval signals with free decay (blue area) and the $\pi-$wait$-\pi$ protocol (red area) for a storage time $t_s=4~\mu$s. 

In contrast to the theoretical prediction of Fig.~\ref{Fig.1}(b), the retrieval photon counts in Fig.~\ref{Fig.3}(a) tend to decrease fast with longer $t_s$, which is attributed to the decrease of optical depth~(OD). To examine this, we measured the OD of the cold gas, shown with blue triangles in Fig.~\ref{Fig.3}(b), displaying the rapid decrease with longer $t_s$. 
Considering that the atomic density distribution, the atomic velocity distribution and the beam profiles are all Gaussian, we can fit the OD with a Gaussian function~\cite{PENG20167507}, shown as the solid line in Fig.~\ref{Fig.3}(b). 
The drop of OD may be due to that the atoms get lost during the long storage time, i.e., a large portion of the atoms in the cloud expand to regions that are no longer covered by the retrieval coupling laser. Armed with this understanding, we analyzed the influence of atom loss on the retrieval efficiency [see section V of the SM], and found that if we correct for the loss of OD, the retrieval efficiency would be much larger, as shown by the solid squares in Fig.~\ref{Fig.3}(a). The fitting to the calibrated data with $A_{3}+B_{3}\exp{-(t_s/\tau_{3})}$ yields $\tau_3= 37.08~\mu$s. 
Besides atom loss, laser noise also hampers the applicability of the protocol [see section IV of the SM].

\begin{figure}[htbp]
    \centering
    \includegraphics[width=\linewidth]{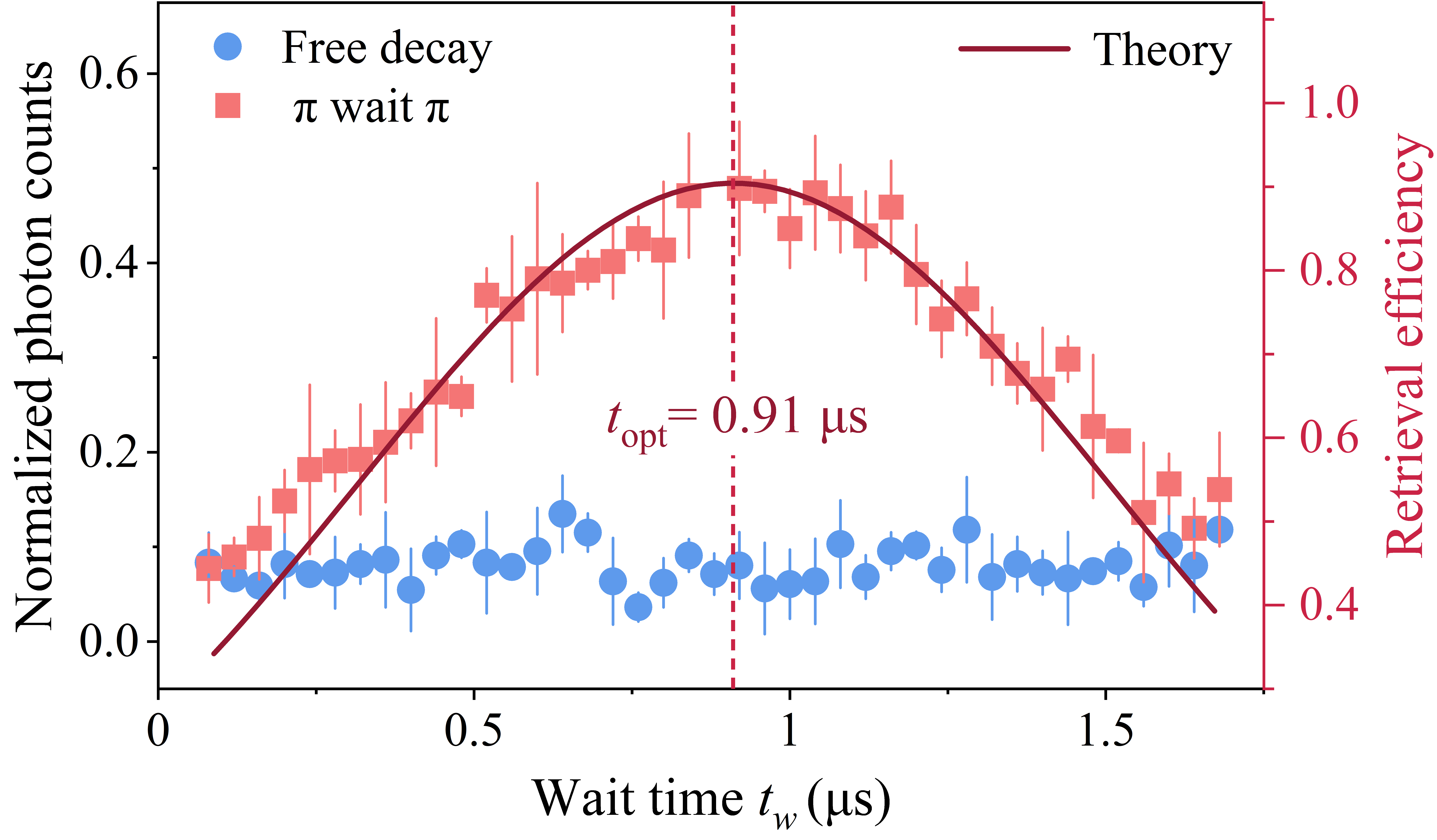}
    \caption{Normalized retrieval photon counts as a function of wait time $t_w$ with fixed $t_s$ = 7~$\mu$s for $\pi-$wait$-\pi$ protocol (squares) and free 
    decay (circles). The error bars show the standard deviation of three independent measurements. The solid curve presents the theoretical retrieval efficiency which indicates the existence of an optimal wait time $t_{\rm opt} = 0.91~\mu$s.}
    \label{Fig.4}
\end{figure}

The 
suppression of the dephasing of RWS hinges on that the wait time should be equal to the optimal value $t_{\rm opt}$. To understand this, we examine the number of the retrieval photons as a function of $t_\emph{w}$ with a fixed storage time $t_s=7~\mu$s, shown by the squares in Fig.~\ref{Fig.4}. Using the $\pi-$wait$-\pi$ protocol, the retrieval efficiency can still be large with moderate deviation of $t_\emph{w}$ from its optimal value $t_{\rm opt}$. As a reference, we also present the data without using our protocol. The solid curve is the theoretical retrieval efficiency [see section III of the SM] by varying $t_\emph{w}$ around $t_{\rm opt}$, which shows that the efficiency should be maximal when $t_\emph{w} = t_{\rm opt}$. The agreement between experiment and theory indicates that the observed coherence enhancement of RWS indeed follows the physical picture outlined in Fig.~\ref{Fig.1}.

Besides the correct choice of the wait time, an equally important factor in our method is the directions of the rephasing lasers. 
In Fig.~S3 of the SM, we present the verification of the directions of the rephasing laser by reversing the directions of the two rephasing laser beams.



 

In summary, we have experimentally demonstrated a state mapping protocol that cancels the motional dephasing of a quantum superposition state. 
Our theory shows that for no atom loss, and minimal laser or electromagnetic noise, and if $\Omega_r\gg k_r v$, the motional dephasing will be completely suppressed.
In this case, Rydberg-state decay remains as the only error channel. The experimental results show an enhancement of the coherence time by one to two orders of magnitude depending on the storage time. Nevertheless, the fact that with an experimental setup highly burdened by laser noise and potential noise of residual fields, the novel protocol can still significantly enhance the coherence of the W state shows that it is possible to use state mapping to quench thermal-motion induced dephasing in atomic and molecular systems. The demonstrated long-lived Rydberg excitation in cold atomic gas shows the feasibility to entangle two single photons separated by tens of microns via weak van der Waals interactions between RWS~\cite{Saffman2010,Adams2020,Shi2021qst}, and thereby provides an alternative route toward quantum networks based on single photons.

This work was supported by the National Natural Science Foundation of China (No. 62175136, 12241408, 12120101004, U2341211 and 12074300); Innovation Program for Quantum Science and Technology (No. 2021ZD0302100, 2023ZD0300902); Fundamental Research Program of Shanxi Province (No. 202303021224007). C.S.A. acknowledges the financial support provided by the UKRI, EPSRC grant reference number EP/V030280/1 (“Quantum optics using Rydberg polaritons”). X.F.S. proposed the protocols with fruitful discussions with Li You, Yan Lu, and T. A. B. Kennedy.

\bibliography{main}

\end{document}